\documentclass[a4paper]{jpconf}
\usepackage{graphicx}
\begin{document}

\title{Magnetic structure of the noncentrosymmetric heavy-fermion superconductor CePt$_3$Si}

\author{ B F{\aa}k$^1$, M Enderle$^1$ and G Lapertot$^2$}

\address{$^1$Institut Laue-Langevin, CS 20156, F-38042 Grenoble Cedex 9, France}

\address{$^2$Universit\'e Grenoble-Alpes and Commissariat \`a l'Energie Atomique, INAC-Pheliqs, F-38000 Grenoble, France}

\ead{fak@ill.fr}

\begin{abstract}
We have determined the magnetic structure of the noncentrosymmetric heavy-fermion superconductor CePt$_3$Si using elastic polarized neutron scattering with field-projected (longitudinal) XYZ polarization analysis. 
We find that the magnetic moment has components both in the tetragonal basal plane and along the tetragonal four-fold axis. 
Such a magnetic structure is not symmetry-allowed if the magnetic phase transition is second order, and may imply that CePt$_3$Si has lower symmetry in the superconducting state than originally thought. 
 \end{abstract}

The tetragonal heavy-fermion compound CePt$_3$Si has received strong attention in recent years because it shows unconventional superconductivity despite its noncentrosymmetric crystal structure \cite{Bauer04,Bauer07,Fujimoto07rev}. 
This lack of inversion symmetry leads to an antisymmetric spin-orbit coupling of Rashba type, 
which mixes spin-singlet and spin-triplet Cooper-pairing channels \cite{Gorkov01}.  
The superconducting transition occurs at $T_c=0.75$~K, 
well inside the magnetically ordered state that sets in at $T_N=2.2$~K. 
This suggests that magnetism may play an important role in the pairing mechanism, 
further highlighted  by the fact that the Cooper pairs are formed by heavy quasiparticles \cite{Bauer04}. 

The magnetic excitations in CePt$_3$Si are well understood. 
The $J=5/2$ multiplet of the $4f^1$ Ce$^{3+}$ ion is split by the tetragonal point symmetry into three doublets.  
The doublet ground state  is $0.46|\pm 5/2\rangle + 0.89 |\mp 3/2 \rangle$, 
as shown by a study combining polarized inelastic neutron scattering 
with soft x-ray absorption spectroscopy  \cite{Willers09}. 
The main exchange interactions have been determined from the dispersion of the somewhat damped spin-wave excitations in the magnetically ordered phase by unpolarized inelastic neutron scattering (INS) \cite{Fak08}. 
Above $T_N$, Kondo-type spin fluctuations with an anisotropic wave-vector dependence are observed by INS \cite{Fak08}. 
As a consequence of the antisymmetric spin-orbit coupling, the spin susceptibility is anomalous with an orthorhombic anisotropy, as shown by polarized INS \cite{Fak14}. 
No resonance or any change in the magnetic excitation spectrum is reported in the superconducting state, despite intense search \cite{Fak08,Inosov11}. 

The magnetic structure of CePt$_3$Si is characterized by a propagation vector ${\bf k}=(0,0,1/2)$, 
but the moment direction is not known. 
Neutron scattering measurements show magnetic Bragg peak intensity for  
wavevector transfers ${\bf Q}\parallel {\bf k}$, 
which implies that the ordered moment is not parallel to the tetragonal $c$ axis. 
It was therefore assumed that the magnetic structure consisted of sheets of collinear moments aligned along an unknown axis in the $a$--$b$ plane, 
with the sheets stacked antiferromagnetically along the $c$ axis  \cite{Metoki04}. 
Based on this assumption, 
the magnitude of the ordered moment was determined to 0.16$\pm$0.01~$\mu_B$ \cite{Fak08,Metoki04}, substantially reduced from the 1.72~$\mu_B$ that can be calculated from the ground-state doublet \cite{Willers09}. 
The hypothesis of the moments lying in the tetragonal basal plane received further support from representation analysis, which showed that for a second-order magnetic phase transition with ${\bf k}=(0,0,1/2)$ in the CePt$_3$Si space group $P4mm$ (No.~99),
the moments are either in the basal plane or along the $c$ axis \cite{Fak08}. 
Very surprisingly, the intensities of the magnetic Bragg peaks were found to increase substantially on the application of a magnetic field {\bf B} along the [001] axis \cite{Kaneko14}. 
On the assumption of moments aligned in the basal plane, 
this suggested that the moment magnitude would double for a field of only 6.6~T along the $c$ axis. 

Because of the small moment, all magnetic structure determinations in CePt$_3$Si are based on elastic neutron scattering measurements on triple-axis spectrometers (TAS), 
a technique that is not very precise in determining moment directions, 
since it is difficult to normalize the intensities measured on different Bragg peaks. 
We have therefore redetermined the magnetic structure using elastic polarized neutron scattering with field-projected (longitudinal) XYZ polarization analysis on the IN20 triple-axis spectrometer. 
The use of polarized neutrons allows to eliminate the $Q$-dependent instrumental normalization factors that make unpolarized TAS measurements problematic. 
We used a standard set-up with Heusler monochromator and analyzer and a Helmholtz coil at the sample position. 
To minimize contributions of unpolarized second-order neutrons from the monochromator, 
we used an incoming neutron wavevector of $k_i=2.57$~\AA$^{-1}$\ and two pyrolytic graphite filters. 
The sample was the same image-furnace grown single crystal of mass 6~g used in many of the previous measurements \cite{Willers09,Fak08,Fak14}, 
mounted with the $a$ and $c$ axes in the scattering plane 
(the lattice parameter are $a=4.072$ and $c=5.442$~\AA). 
The flipping ratio on nuclear Bragg peaks exceeded 23, which makes polarization corrections unnecessary. 

For the determination of the magnetic structure, we measured thirteen magnetic Bragg peaks, 
namely  
$(0,0,1/2)$, $(0,0,3/2)$, $(0,0,5/2)$, 
$(1,0,\pm 1/2)$, $(1,0,\pm 3/2)$, $(1,0,\pm 5/2)$, 
$(2,0,\pm 1/2)$, and $(2,0,\pm 3/2)$
with seven different polarized neutron scattering cross-sections $\alpha\beta$: 
$xx$, $x\bar{x}$, $\bar{x}x$, $yy$, $y\bar{y}$, $zz$, and $z\bar{z}$,
where $\alpha$ and $\beta$ are the incoming and scattered neutron polarization, respectively, 
the  directions $x$, $y$, $z$ correspond to the neutron polarization being along {\bf Q}, perpendicular to {\bf Q} in the scattering plane, and perpendicular to the scattering plane, respectively, 
and $\bar{x}$ means antiparallel to $x$. 

Since neutron scattering observes only the component of the magnetic moment {\bf M} that is perpendicular to {\bf Q}, here called ${\bf M}_\perp$, 
it is straightforward to show that the measured cross-sections $I_{\alpha\beta}$ are related to the $y$ and $z$ projected moment components $M_{\perp y}$ and $M_{\perp z}$ via
\begin{eqnarray}
 I_{x\bar x} - I_{y\bar y} &=&  
 |M_{\perp,y}|^2 - 2 {\rm Im} \{ M_{\perp,y}^\star M_{\perp,z} \}  \label{sfxy} \\
I_{yy} - I_{xx} &=& |M_{\perp,y}|^2  \label{nfxy} \\
 I_{x\bar x} - I_{z\bar z} &=& 
 |M_{\perp,z}|^2 - 2 {\rm Im} \{ M_{\perp,y}^\star M_{\perp,z} \} \label{sfxz} \\
I_{zz} - I_{xx} &=& |M_{\perp,z}|^2 \label{nfxz}\\
I_{x\bar x} - I_{\bar xx} &=&   
-4 {\rm Im} \{ M_{\perp,y}^\star M_{\perp,z} \}  \label{sfxx}
 .
\end{eqnarray}
The intensities $I_{\alpha\beta}$ were normalized to the beam monitor, and the background obtained from measurements with the sample rotated away from the Bragg position by $2^\circ$ or in the paramagnetic phase at $T=7$~K was subtracted.  
The chiral term ${\rm Im} \{ M_{\perp,y}^\star M_{\perp,z} \}$ obtained from (\ref{sfxx}) 
is zero within the precision of the present measurements. 
This is expected, since the propagation vector of CePt$_3$Si does not lead to any chirality. 

In the actual geometry of the experiment, with $\theta$ denoting the angle between {\bf Q} and the tetragonal basal plane, the magnetic moment components along the crystallographic axes, 
$M_a$, $M_b$, and $M_c$, are related to $M_{\perp,y}$ and $M_{\perp,z}$ via 
\begin{equation}
|M_{\perp z}|^2 = |M_b|^2 
\label{EqMz}
\end{equation}
and 
\begin{eqnarray}
|M_{\perp y}|^2 &=& |M_a|^2 \sin^2\theta + |M_c|^2\cos^2\theta 
- 2 {\rm Re} \{ M_a^\star M_c \} \sin\theta\cos\theta =
\nonumber\\ &=&
|M_a|^2 \sin^2\theta + |M_c|^2\cos^2\theta  -2M_aM_c\sin\theta\cos\theta , 
\label{EqMy}
\end{eqnarray}
where in the last equality we have assumed real (Fourier components of the) magnetic moments, which is justified in the case of a ${\bf k}=(0,0,1/2)$ propagation vector. 
By measuring $|M_{\perp y}|^2$ for several inequivalent values of {\bf Q}, 
and hence several values of $\theta$, both $M_a$ and $M_c$ can be extracted from (\ref{EqMy}). 

We introduce the notation $|M_{\perp y}|^2_{\pm l}$ for measurements performed at 
${\bf Q}=(h,0,\pm l/2)$ with $h>0$. 
The mixed $M_aM_c$ term can be estimated from the difference 
\begin{equation}
|M_{\perp y}|^2_{+l} - |M_{\perp y}|^2_{-l}  
= -2M_aM_c \sin(2\theta) .
\label{EqMac}
\end{equation}
Within the precision of the measurements, $M_aM_c=0$ (see table \ref{TableResults}). 
This mixed term can also be eliminated from (\ref{EqMy}), by taking the average of 
$|M_{\perp y}|^2_{+l}$ and $|M_{\perp y}|^2_{-l}$. 

\begin{table}[!b]
\caption{\label{TableResults} Moment ratios of in-plane $|M_a/M_b|^2$, 
out-of-plane $|M_c/M_b|^2$, and mixed $M_aM_c/|M_b|^2$ contributions 
from spin-flip and non-spin-flip cross-sections, and their average. 
Statistical errors (one standard deviation) are given in parenthesis
and observed standard deviations are given as $\pm$ values.}
\begin{center}
\begin{tabular}{llll}
\br
 Moment ratio & Spin-Flip & Non-Spin-Flip & Average \\ 
\mr
$|M_a/M_b|^2$ & 0.68(4)$\pm$0.03 &  0.62(4)$\pm$0.05 &  0.65(3)$\pm$0.05\\
$|M_c/M_b|^2$ & 3.2(2)$\pm$1.2  & 3.2(2)$\pm$0.8 &  3.2(1)$\pm$1.0\\
$M_aM_c/|M_b|^2$ &  0.08$\pm$0.12 &  0.06$\pm$0.11 &  0.07$\pm$0.11\\
\br
\end{tabular}
\end{center}
\end{table}

The ratio of the in-plane components of the magnetic moment are obtained 
for reflections with $h=0$ ($\theta=\pi/2$), 
\begin{equation}
\alpha^2 \equiv \frac { |M_{\perp y}|^2 }  { |M_{\perp z}|^2 }_{h=0} = \frac { |M_a|^2 } {   |M_b|^2 } .
\label{EqAlpha}
\end{equation}
Averaging over the measured magnetic Bragg reflections with $h=0$, 
we find $\bar\alpha^2 = 0.65 \pm 0.05$ for both spin-flip and non-spin-flip cross-sections, 
and hence $\bar\alpha=|M_a|/|M_b|=0.81 \pm 0.03$ (see table \ref{TableResults}). 
This implies that the population of in-plane S domains is unequal, 
and also that the in-plane component of the moment is not along a [110]-type axis. 
In other words, the basal-plane projected moment is either in a general direction with four in-plane S domains or along a [100]-type axis with two unequally populated in-plane S domains. 

The out-of-plane moment ratio is obtained from (\ref{EqMz}) and (\ref{EqMy}) after averaging over 
$|M_{\perp y}|^2_{+l}$ and $|M_{\perp y}|^2_{-l}$ for $h\neq 0$ Bragg peaks, as discussed, 
\begin{equation}
\frac {|M_c|^2}{ |M_b|^2} =  \frac{1}{\cos^2\theta} \, \frac {|M_{\perp y}|^2}{ |M_{\perp z}|^2}_{h\neq 0} 
- \bar\alpha^2 \tan^2\theta ,
\label{EqGamma}
\end{equation}
using the value of $\bar\alpha$ determined above. 
Note that by normalizing the data to $|M_b|^2 = |M_{\perp z}|^2$, 
absorption corrections as well as Lorentz and magnetic form factors are factored out.  
We find that $|M_c/M_{ab}|^2=3.2 \pm 1.0$
(or $|M_c/M_{ab}|=1.8 \pm 0.3$, see table \ref{TableResults}) 
for both spin-flip and non-spin-flip cross-sections. 
This implies that the moment component along the tetragonal $c$ axis is {\it substantial}, 
in contrast to assumptions hitherto. 

The observation that the ordered moment direction is neither in the basal plane nor along the tetragonal axis has two important implications. 
One is that the field-induced intensity increase of magnetic Bragg peaks 
observed by Kaneko {\it et al.} \cite{Kaneko14} 
cannot be directly related to the size of the magnetic moment, 
since a change in the domain population or a reorientation of the magnetic moment in applied field would also change the Bragg peak intensity. 
For a magnetic field in the [001] direction, 
a  domain repopulation is not likely since all S domains have the same energy. 
However, the Zeeman energy due to the magnetic field will compete with the anisotropy that produces the initial tilt angle of the moment direction, and the total energy of the system will be lowered if the magnetic moments turn further towards the basal plane. Such a tilting will increase the intensity of the (0,0,1/2) Bragg peak, as observed experimentally \cite{Kaneko14}.

The second implication is that the magnetic order established at $T_N$ leads to a lower symmetry than expected for a second order phase transition. 
This means that the phase transition is either (weakly) first order or that 
there are two phase transitions close to each other near $T_N$. 
Both scenarios lead to a lowering of the symmetry in the magnetically ordered phase, 
and hence in the superconducting phase, 
with important consequences for the understanding of the superconducting mechanism.

\end{document}